\RequirePackage{fix-cm}
\documentclass[onecolumn,epjc3]{svjour3}  
\usepackage{amsmath, amssymb, amsfonts}
\smartqed  
\RequirePackage{graphicx}

\journalname{Eur. Phys. J. C}

\begin{document}

\title{On the initial spin periods of magnetars born in weak supernova explosions and their gravitational wave radiation
}
\subtitle{}


\author{Yu-Long Yan\thanksref{addr1}
        \and
        Quan Cheng\thanksref{e,addr1} 
        \and
        Xiao-Ping Zheng\thanksref{addr1} 
        \and Xia-Xia Ouyang\thanksref{addr1}
}

\thankstext{e}{e-mail: qcheng@ccnu.edu.cn}


\institute{Institute of Astrophysics, Central China Normal
University, Wuhan 430079, China \label{addr1}}

\date{Received: date / Accepted: date}

\maketitle

\begin{abstract}
The initial spin periods of newborn magnetars are \textbf{strongly associated with the origin of their strong magnetic fields, both of which can affect the electromagnetic radiation and gravitational waves (GWs) emitted at their birth.} Combining the upper limit $E_{\rm SNR}\lesssim10^{51}$ erg on the explosion energies of \textbf{the supernova (SN) remnants around slowly-spinning magnetars} with a detailed investigation on the evolution of newborn magnetars, we set constraints on the initial spin periods of magnetars \textbf{born in weak SN explosions}. Depending on the conversion efficiency $\eta$ of the electromagnetic energy of \textbf{these} newborn magnetars into the kinetic energy of SN ejecta, the minimum initial spin periods of \textbf{these} newborn magnetars are $P_{\rm i, min}\simeq 5-6$ ms for an ideal efficiency $\eta=1$, $P_{\rm i, min}\simeq 3-4$ ms for a possible efficiency $\eta=0.4$, and $P_{\rm i, min}\simeq 1-2$ ms for a relatively low efficiency $\eta=0.1$. \textbf{Based on these constraints and adopting reasonable values for the physical parameters of the newborn magnetars, we find that their GW radiation at $\nu_{\rm e,1}=\nu$ may be undetectable by the Einstein Telescope (ET) since the maximum signal-to-noise ratio (${\rm S/N}$) is only 2.41 even the sources are located at a very close distance of 5 Mpc, where $\nu$ are the spin frequencies of the magnetars. At such a distance, the GWs emitted at $\nu_{\rm e,2}=2\nu$ from the newborn magnetars with dipole fields $B_{\rm d}=5\times10^{14}$ and $10^{15}$ G may be detectable by the ET because ${\rm S/N}$ are 10.01 and 19.85, respectively. However, if these newborn magnetars are located at $20$ Mpc away in the Virgo supercluster, no GWs could be detected by the ET due to low ${\rm S/N}$.} 

\end{abstract}

\section{Introduction}\label{Sec I} 
Magnetars are a subclass of neutron stars (NSs) that behave as soft gamma-ray repeaters (SGRs) and anomalous X-ray pulsars (AXPs) in observations and have dipole magnetic fields with a typical strength of $\sim10^{14}-10^{15}$ G \cite{Olausen:2014} and stronger multipolar magnetic fields in the exterior \cite{Tiengo:2013}. In the interior of magnetars, it is generally considered that toroidal magnetic fields with strengths of at least comparable to or even much higher than that of the external dipole fields that possibly exist (e.g., \cite{Stella:2005,Braithwaite:2009,Ciolfi:2009,Makishima:2014}). Possible evidence for the existence of stronger toroidal fields in the interior are mainly the periodic pulse-phase modulations in the hard X-ray emissions from the
magnetars 4U 0142+61, 1E 1547-5408, and SGR 1900+14 \cite{Makishima:2014,Makishima:2021a,Makishima:2021b}, and the 2004 December 27 giant flare from SGR 1806-20 \cite{Stella:2005}. The traditional magnetars spin not so fast as most of the ordinary radio pulsars and have typical spin periods of $\sim1-12$ s \cite{Olausen:2014}\footnote{See the McGill online magnetar catalog:
https://www.physics.mcgill.ca/$\sim$pulsar/magnetar/main.html},
though rather slow rotations with a period of 6.67 hr for the magnetar located in the supernova (SN) remnant RCW 103 \cite{De Luca:2006,Daia:2016}, and a possible period of 1091 s for the magnetar GLEAM-X J162759.5-523504.3 were reported in the literature \cite{Hurley-Walker:2022}. Another type of magnetars dubbed newborn millisecond magnetars is generally suggested to have millisecond spin periods and dipole fields of $\sim10^{14}-10^{15}$ G. These magnetars are thought to be produced in the core collapse
of massive stars and the merger of binary NSs (see, e.g., \cite{Zhang:2001,Rowlinson:2013,Gao:2016,Metzger:2018}). Their fast spin and strong magnetic fields render them possible central engines of long/short gamma-ray bursts (GRBs) \cite{Zhang:2001,Rowlinson:2013,Dai:1998,Metzger:2008,Dall'Osso:2011}, superluminous supernovae (SLSNe) \cite{Kasen:2010,Yu:2017}, fast radio bursts \cite{Zhang:2014,Metzger:2017}, and fast blue
optical transients \cite{Yu:2015,Liu:2022}. Although the spin periods of the two types of magnetars are quite different, their dipole fields are generally the same, and the strong magnetic fields may be related to the fast spin of NSs.

\textbf{The origin of magnetars' strong magnetic fields} has garnered particular attention since \textbf{their existence was first proposed} \cite{Duncan:1992,Thompson:1993}. Based on the specific amplification mechanisms of NS magnetic fields, the origins of strong magnetic fields can generally be divided into two kinds, namely the dynamo and the fossil origins. The dynamo origins mainly include the $\alpha-\Omega$ and convective dynamos that act in millisecond protoneutron stars \cite{Duncan:1992,Raynaud:2020}, dynamo processes that arise from the Kelvin-Helmholtz \cite{Price:2006} or magnetorotational instability \cite{Akiyama:2003} in nascent millisecond NSs, Tayler-Spruit dynamo in protoneutron stars which were spun up to millisecond periods through the fall-back accretion \cite{Barrere:2022}, dynamo due to r-mode and Tayler instabilities in newborn millisecond NSs \cite{Cheng:2014}. Therefore, fast spin is indispensable for the arising of these dynamo processes, and after the end of the dynamo processes, the newly formed magnetars will probably have millisecond spin periods. On the other hand, the fossil origin refers to the magnetic flux conservation during the core collapse of highly magnetized massive stars, which could also lead to the formation of strong magnetic fields \cite{Ferrario:2006}, and the magnetars formed in this way may not have initial spin periods of milliseconds.

Generally, if the magnetars formed in the core collapse of massive stars have millisecond initial spin periods, huge spin energy of the magnetars could be extracted through magnetic dipole (MD) radiation and relativistic particle wind and injected into SN ejecta. For instance, the spin energy of a newborn magnetar with an initial spin period of $\sim1$ ms can be as large as $\sim10^{52}$ erg \cite{Duncan:1992}. \textbf{Such huge energy seems to be in contradiction with the explosion energies of the remnants around some magnetars formed in weak SNe, which are derived to be $\lesssim10^{51}$ erg by analyzing the X-ray spectra of these remnants \cite{Vink:2006,Martin:2014,Zhou:2019}.} This indicates that initial spin periods \textbf{of at least some magnetars may be much larger than $\sim1$ ms, in support of} the idea that strong magnetic fields of these magnetars are produced due to magnetic flux conservation (e.g., \cite{Vink:2006,Zhou:2019}).

In fact, the dynamo origin of strong magnetic fields of magnetars still may not be excluded. Such a conclusion was reached by comprehensively investigating the spin, magnetic tilt angle, and thermal evolutions of newborn magnetars as that conducted in \cite{Dall'Osso:2009}, which showed that a large proportion of the spin energy of a newborn magnetar can be lost via gravitational wave (GW) radiation without powering the SN ejecta even if the initial spin is $\sim1-2$ ms. The requirement is that the magnetar should have external dipole and internal toroidal fields of appropriate strengths and the latter should be stronger than the former \cite{Dall'Osso:2009}. The tilt angle of this newborn millisecond magnetar may increase very quickly to a relatively large value that is beneficial for GW radiation because the free-body precession of this magnetar could be quickly damped due to the bulk viscosity of stellar matter \cite{Dall'Osso:2009}. However, as indicated by some dynamo processes (e.g., Refs. \cite{Duncan:1992,Raynaud:2020,Cheng:2014}), and the stability of internal magnetic field configuration \cite{Braithwaite:2009,Ciolfi:2009,Lander:2009}, the strength ratio of toroidal to dipole fields may distribute in a relatively wide range of $\sim2-100$. Such a range for the strength ratio was also verified by observations of the giant flare from SGR 1806-20 \cite{Stella:2005}, periodic modulations in the hard X-ray emissions of magnetars 4U 0142+61, 1E 1547-5408, and SGR 1900+14 \cite{Makishima:2014,Makishima:2021a,Makishima:2021b}, X-ray afterglows of some GRBs \cite{Gao:2016,Lin:2022}, and lightcurves of SLSNe \cite{Moriya:2016}. When the newborn magnetar's toroidal field is much stronger than the dipole field, its tilt angle could not increase to a large value that benefits GW radiation on a short timescale by virtue of bulk viscosity \cite{Dall'Osso:2009,Cheng:2015}. In this case, the tilt angle would increase to a large value only if the stellar free-body precession is damped by the stronger viscosity resulting from the scattering of the relativistic electrons off superfluid neutrons in the NS core \cite{Dall'Osso:2009,Cheng:2015,Alpar:1988,Cutler:2002}. While the electrons in the core follow the instantaneous rotation of the crust, the superfluid neutrons cannot, thus the viscosity is also considered to arise from core-crust coupling \cite{Dall'Osso:2009,Cheng:2015}. It appears only when the NS has cooled down so that a solid crust has been formed and the neutrons in the core have become superfluid \cite{Dall'Osso:2009,Cheng:2015,Cheng:2018}. Consequently, when the magnetar has quite strong toroidal field, rather than only considering the bulk viscosity of stellar matter as that done in \cite{Dall'Osso:2009}, the viscosity due to core-crust coupling should also be involved in the study of tilt angle evolution. Based on the more detailed investigation of tilt angle evolution, by using the upper limit $E_{\rm SNR}\lesssim10^{51}$ erg on the explosion energies of the SN remnants around the \textbf{slowly-spinning magnetars formed in weak SN explosions}, we set new constraints on the initial spin periods of these magnetars.

\textbf{The results show that depending on the conversion efficiency $\eta$ of electromagnetic (EM) energy supplied by the newborn magnetars into kinetic energy of the SN ejecta, the minimum initial spin periods of magnetars are $P_{\rm i, min}\simeq 1-2$ ms for a relatively low efficiency $\eta=0.1$. However, we have $P_{\rm i, min}\simeq 3-4$ ms, and $5-6$ ms when a possible efficiency $\eta=0.4$ \cite{Zhou:2019,Woosley:2010}, and an ideal efficiency $\eta=1$ are assumed, respectively.} The resultant $P_{\rm i, min}$ of the newborn magnetars do not vary
significantly with the changes of both dipole and toroidal magnetic fields. Our constraints on the initial spin periods of the newborn magnetars \textbf{differ by a factor of three} from that of \cite{Dall'Osso:2009}, which showed that the initial spin periods can be $\sim1-2$ ms when $\eta=1$ is adopted. 

The content of this work is organized as follows. In Sec. \ref{Sec II}, we show the evolution of newborn magnetars. \textbf{We constrain the initial spin periods of slowly-spinning magnetars formed in weak SN explosions in Sec. \ref{Sec III}. Based on the constraints, the GW radiation from these newborn magnetars are investigated in Sec. \ref{Sec IV}.} Finally, we give the conclusion and some discussions in Sec. \ref{Sec V}.

\section{Evolution of newborn magnetars}\label{Sec II}

\textbf{After the core collapse of a massive star, a strong internal toroidal magnetic field with volume-averaged strength $\bar{B}_{\rm t}$ and a surface dipole magnetic field $B_{\rm d}$ may be formed in a newborn millisecond magnetar due to some dynamo processes \cite{Duncan:1992,Raynaud:2020,Akiyama:2003,Cheng:2014}. Though a stable twisted-torus magnetic configuration consisting of both poloidal and toroidal magnetic fields may be produced in the magnetar interior \cite{Braithwaite:2004}, $\bar{B}_{\rm t}$ could still play a dominant role \cite{Stella:2005,Braithwaite:2009}. With the presence of strong $\bar{B}_{\rm t}$, the magnetar would deform into a prolate ellipsoid and emit GWs. Moreover, because of the fast spin of the newborn magnetar, r-mode instability could arise and represent another way of emitting GWs \cite{Andersson:1998}. The newborn magnetar's strong $B_{\rm d}$ makes MD radiation an effective torque that can spin down the star. It is also possible that during the first few minutes shortly after the birth of the magnetar, strongly magnetized, relativistic neutrino-driven wind may emerge, leading to the loss of stellar angular momentum \cite{Thompson:2004}. Nevertheless, previous work seems to favor a quite small saturation amplitude of r-mode in NSs \cite{Mahmoodifar:2013} and an ineffective braking torque caused by the neutrino-driven wind in comparison with the MD radiation from the newborn magnetar \cite{Sur:2021}, hence losses of stellar angular momentum due to the two mechanisms could be neglected. In this work, we consider that the newborn magnetar spins down mainly due to MD and magnetically deformed GW radiation, the magnetar's spin evolution thus has the following form \cite{Cutler:2000,Spitkovsky:2006}:}
\begin{eqnarray}\label{domega}
    \dot{\Omega}=-\frac{B_{\rm d}^2R^6\Omega^3}{6Ic^3}(1+\sin^2\chi)-\frac{2G\epsilon_{\rm B}^2I\Omega^5}{5c^5}\sin^2\chi(1+15\sin^2\chi),
\end{eqnarray}
where $\Omega$, $R$, and $\chi$ are respectively the angular velocity, radius, and magnetic tilt angle (the angle between the magnetic and spin axes) of the newborn magnetar and $I=0.35MR^2$ the stellar moment of inertia with $M$ representing the mass of the magnetar \cite{Lattimer:2001}. In the case of toroidal-dominated internal fields, the ellipticity of magnetic deformation is $\epsilon_{\rm B}=-5\bar{B}_{\rm t}^2R^4 /(6GM^2)$ \cite{Cutler:2002}.

Initially, $\chi$ of the newborn magnetar may be very tiny as inferred from the dynamo processes \cite{Dall'Osso:2009}. Misalignment of the stellar magnetic and spin axes will lead to the free-body precession of the magnetar's magnetic axis around the spin axis. For the newborn magnetar with toroidal-dominated internal fields, viscous dissipation of the star's precessional energy will result in an anti-aligned torque between the two axes, thus increasing $\chi$ \cite{Dall'Osso:2009}. Meanwhile, the MD and GW radiation can give rise to aligned torques, thus decreasing $\chi$ \cite{Dall'Osso:2009,Cutler:2000,Philippov:2014}. Depending on the specific mechanisms that lead to the dissipation of precessional energy, the evolution of $\chi$ can be roughly divided into two stages \cite{Dall'Osso:2009,Cheng:2015,Cheng:2018}. The first stage starts from the formation of the uniformly rotating newborn magnetar and ends at the time when the magnetar has sufficiently cooled down so that its solid crust has formed and the neutrons in its core have become superfluid \cite{Dall'Osso:2009,Cheng:2015}. During this stage, the stellar temperature is extremely high ($\sim10^{10}$ K), hence the deformed magnetar is a liquid ellipsoid. Damping of the newborn magnetar's free-body precession comes from the bulk viscosity of the liquid dense matter \cite{Dall'Osso:2009}. Consequently, the evolution of $\chi$ in this stage is determined by the competition between damping of the free-body precession due to bulk viscosity and MD and GW radiation, which can be written as \cite{Dall'Osso:2009,Cheng:2018}
\begin{eqnarray}\label{dchi}
    \dot{\chi}=\frac{\cos\chi}{\tau_{\rm d}\sin\chi}-\frac{2G}{5c^5}I\epsilon_{\rm B}^2\Omega^4\sin\chi\cos\chi\left(15\sin^2\chi+1\right)-\frac{B_{\rm d}^2R^6\Omega^2}{6Ic^3}\sin\chi\cos\chi,
\end{eqnarray}
where $\tau_{\rm d}$ represents the damping timescale of free-body precession. When bulk viscosity of dense matter \textbf{plays a dominant role}, $\tau_{\rm d}$ has the following form \cite{Dall'Osso:2009,Cheng:2018}
\begin{eqnarray}\label{taud1}
    \tau_{\rm d}=\tau_{\rm bv}\simeq 3.9~{\rm s}\frac{\cot^2\chi}{1+3\cos^2\chi}\left(\frac{\bar{B}_{\rm t}}{10^{16}~{\rm G}}\right)^2\left(\frac{P}{1~{\rm ms}}\right)^2\left(\frac{T}{10^{10}~{\rm K}}\right)^{-6},
\end{eqnarray}
where $\tau_{\rm bv}$ is the damping timescale of free-body precession due to bulk viscosity, $P=2\pi/\Omega$ and $T$ are respectively the newborn magnetar's spin period and temperature.

Since free-body precession of the newborn magnetar with toroidal-dominated internal fields is gradually damped by bulk viscosity, in principle, the tilt angle would increase and finally may achieve $\chi=\pi/2$, which just corresponds to the minimum spin energy state of the magnetar \cite{Dall'Osso:2009,Cheng:2018,Cheng:2019}. However, when the toroidal field is large enough, the growth of $\chi$ will be suppressed \cite{Dall'Osso:2009,Cheng:2015}. With the cooling and spin-down of the
newborn magnetar, the effect of bulk viscosity on the damping of stellar free-body precession is weakened. When the magnetar has sufficiently cooled down because of neutrino emission, a solid crust will form in the exterior and the neutrons in the NS core will become superfluid \cite{Page:2004,Chamel:2008,Page:2011}. If the
tilt angle has not yet increased to $\chi=\pi/2$ in the first stage, the second stage of evolution will initiate. In this stage, the viscosity due to core-crust coupling becomes effective and can dissipate the precessional energy of the newborn magnetar on a timescale \cite{Stella:2005,Alpar:1988,Cutler:2002}
\begin{eqnarray}\label{taucc}
    \tau_{\rm cc}\simeq \xi P / \epsilon_{\rm B},
\end{eqnarray}
where $\xi$ is the number of precession cycles. Although in previous work various of methods were suggested to determine the value of $\xi$ (e.g., \cite{Alpar:1988,Cutler:2002,Cheng:2019,Hu:2023,Yan:2024}), its exact value still remains highly uncertain. In this work, we take $\xi=10^4$, which is reasonable if
the core-crust coupling is caused by scattering between superfluid neutrons and relativistic electrons \cite{Alpar:1988,Cutler:2002}. Such a value for $\xi$ is also consistent with the results obtained by using the measured timing data and tilt angles of several young pulsars \cite{Hu:2023,Yan:2024}. In addition to the bulk viscosity, in the second stage, damping of stellar free-body precession due to core-crust coupling can also increase the magnetar's tilt angle until $\chi=\pi/2$ is achieved eventually. Therefore, the evolution of $\chi$ in this stage still follows the form given in Eq. (\ref{dchi}), however, the damping timescale now is determined by
\begin{eqnarray}\label{taud2}
    \frac{1}{\tau_{\rm d}}=\frac{1}{\tau_{\rm bv}}+\frac{1}{\tau_{\rm cc}}.
\end{eqnarray}

One thus can see that both the tilt angle evolution in the first stage (Eqs. (\ref{dchi}) and (\ref{taud1})) and the onset of the second stage depends on the stellar temperature $T$. As generally considered, the temperature at which the solid crust is formed may be close to the critical temperature $T_{\rm c}$ for neutrons in the core to be superfluid in the $^3P_2$ channel \cite{Cheng:2018,Chamel:2008,Beloin:2018}, both are $\sim10^9$ K \cite{Dall'Osso:2009}. For simplicity, we assume that both formation of the solid crust and occurrence of neutron superfluidity in the core is at $T_{\rm c}$ \cite{Cheng:2015,Cheng:2018}, thus the second stage of tilt angle evolution will begin when the newborn magnetar cools down to $T_{\rm c}$. For the newborn magnetar with typical mass $M=1.4M_\odot$ and radius $R=12$ km, it may cool down mainly through modified Urca neutrino processes during the first $10^6$ yrs after its birth \cite{Page:2006}. Assuming that the whole NS is isothermal for simplicity \cite{Cheng:2018}, the stellar temperature evolution can be expressed as
\begin{eqnarray}\label{T}
    C_{\rm V}\frac{d T}{d t}=-L_{\nu,~{\rm MU}},
\end{eqnarray}
where $C_{\rm V}\approx10^{39}T_{9}~{\rm erg}/{\rm K}$ is the NS's total specific heat, $L_{\nu,~{\rm MU}}\approx7\times10^{39}T_{9}^{8}~{\rm erg}/{\rm s}$ the total luminosity of the modified Urca neutrino emission with the notation $T_9=T/10^9~{\rm K}$ adopted \cite{Page:2006}.

\section{Constraints on the initial spin periods of magnetars formed in weak SN explosions}\label{Sec III}

\begin{figure}
\begin{center}
\includegraphics[width=4.6 in, height=4.2 in,
clip]{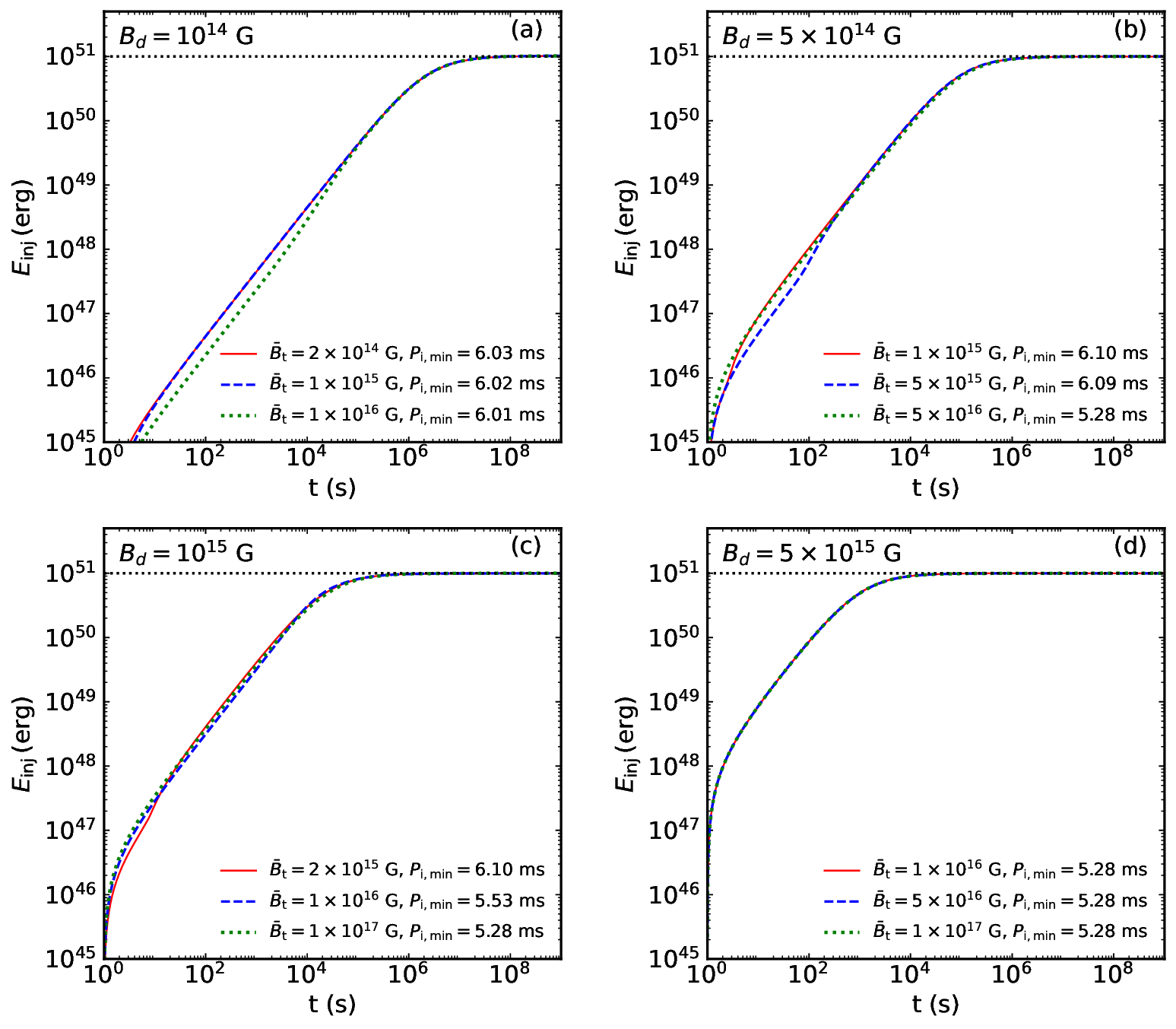} \caption{Evolution of the \textbf{cumulative} energy injected into the SN ejecta by the newborn magnetar $E_{\rm inj}$ with time $t$. In panels (a)-(d), we respectively adopt four typical strengths for the newborn magnetar's dipole field as $B_{\rm d}=10^{14}$, $5\times10^{14}$, $10^{15}$, and $5\times10^{15}$ G. In panels (a)-(c), three values are adopted for the toroidal field as $\bar{B}_{\rm t}=2B_{\rm d}$, $10B_{\rm d}$, and $100B_{\rm d}$, while in panel (d), the maximum value is limited to $\bar{B}_{\rm t}=10^{17}$ G. Assuming an ideal conversion efficiency $\eta=1$, the minimum initial spin period $P_{\rm i, min}$ of the newborn magnetar is obtained for each set of $B_{\rm d}$ and $\bar{B}_{\rm t}$, as indicated in the legends. The black dotted line in each panel
represents the upper limit $E_{\rm SNR}\lesssim10^{51}$ erg on the explosion energies of the SN remnants associated with magnetars \textbf{formed in weak SN explosions}.} \label{Fig1}
\end{center}
\end{figure}

\begin{figure}
\begin{center}
\includegraphics[width=4.6 in, height=4.2 in,
clip]{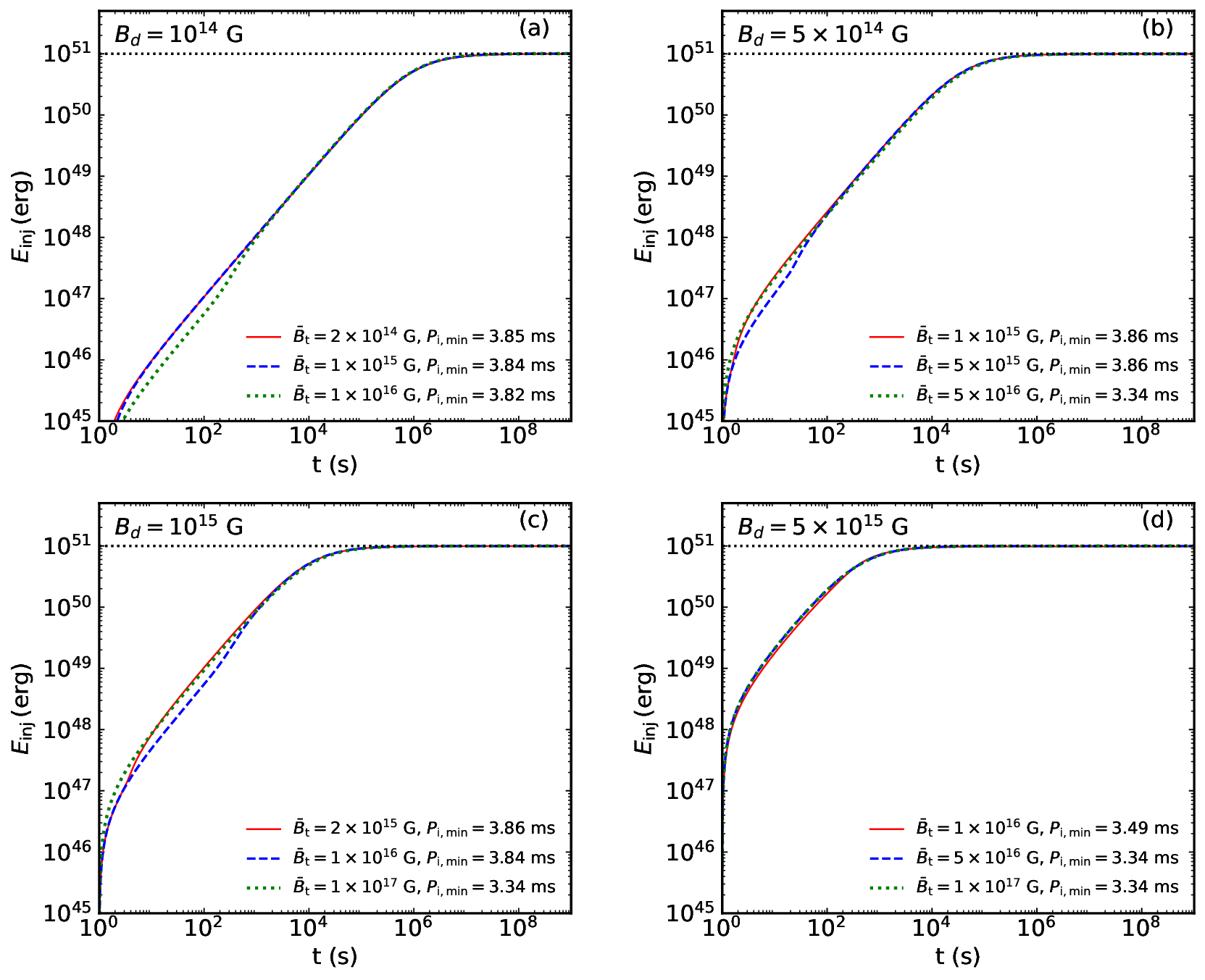} \caption{\textbf{Evolution of the cumulative energy injected into the SN ejecta by the newborn magnetar $E_{\rm inj}$ with time $t$. In panels (a)-(d), we respectively adopt four typical strengths for the newborn magnetar's dipole field as $B_{\rm d}=10^{14}$, $5\times10^{14}$, $10^{15}$, and $5\times10^{15}$ G. In panels (a)-(c), three values are adopted for the toroidal field as $\bar{B}_{\rm t}=2B_{\rm d}$, $10B_{\rm d}$, and $100B_{\rm d}$, while in panel (d), the maximum value is limited to $\bar{B}_{\rm t}=10^{17}$ G. Assuming a possible efficiency $\eta=0.4$, the minimum initial spin period $P_{\rm i, min}$ of the newborn magnetar is obtained for each set of $B_{\rm d}$ and $\bar{B}_{\rm t}$, as indicated in the legends. The black dotted line in each panel represents the upper limit $E_{\rm SNR}\lesssim10^{51}$ erg on the explosion energies of the SN remnants associated with magnetars formed in weak SN explosions.}} \label{Fig2}
\end{center}
\end{figure}

With the spin-down of the newborn magnetar, part of its spin energy can be lost via MD radiation and injected into the ejecta it is embedded in, thus the associated SN remnant may be energized. The \textbf{cumulative} energy injected into the ejecta by the central magnetar can be estimated as
\begin{eqnarray}\label{Einj}
    E_{\rm inj}=\eta\int_0^{t} L_{\rm dip}dt,
\end{eqnarray}
where $L_{\rm dip}=B_{\rm d}^2R^6\Omega^4(1+\sin^2 \chi)/(4c^3)$ is the luminosity of MD radiation of the newborn magnetar \cite{Spitkovsky:2006,Kashiyama:2016}. The injected energy will saturate at $t=t_{\rm sat}$, thus the saturation energy $E_{\rm inj, s}$ is the total energy injected into the ejecta by the newborn magnetar. For the physical parameters of newborn magnetars adopted in this work, we generally have $t_{\rm sat}\lesssim 10^7$ s (as shown in Figs. \ref{Fig1} and \ref{Fig2}), which is much shorter than the ages of the SN remnants of magnetars \cite{Vink:2006,Zhou:2019}. Therefore, energy injection ends soon after the birth of the magnetar. $\eta$ is the conversion efficiency of the EM energy into the kinetic energy of the ejecta. The value of $\eta$ is highly uncertain, though a possible value $\eta=0.4$ was proposed in \cite{Woosley:2010}. \textbf{In our calculations, $\eta=0.1$, $0.4$, $0.7$, and $1.0$ are used.}

The SN remnants associated with magnetars are generally considered to be in the Sedov phase, in which the SN ejecta expands adiabatically in a uniform interstellar medium \cite{Sedov:1959,Ostriker:1988,Truelove:1999}. Based on this assumption, the explosion energies of SN remnants associated with magnetars can be estimated if the radius and velocity of the shock, and the interstellar medium density can be determined from observations (see e.g., \cite{Vink:2006,Zhou:2019}). By performing an analysis of the overall X-ray spectra of SN remnants Kes 73, CTB 109, and N49 (which respectively host magnetars 1E 1841-045, 1E 2259+586, and SGR 0526-66), the above quantities and explosion energies of these remnants were obtained \cite{Vink:2006,Martin:2014}. Following Vink \& Kuiper \cite{Vink:2006}, the explosion energies of Kes 73, CTB 109, and N49 are $E_{\rm SNR}=(0.5\pm0.3)\times10^{51}$, $(0.7\pm0.3)\times10^{51}$, and $(1.3\pm0.3)\times10^{51}$ erg, respectively. Recently, Zhou et al. \cite{Zhou:2019} analyzed the spatially resolved X-ray spectra of the SN remnants Kes 73, N49, and RCW 103 (the host of magnetar 1E 161348-5055) in detail and derived their explosion energies, which are respectively about $5.4\times10^{50}$, $1.7\times10^{51}$, and $1.0\times10^{50}$ erg. We thus can see that though the SN remnant N49 may have explosion energy that slightly surpasses $10^{51}$ erg, the majority of remnants associated with \textbf{these slowly-spinning} magnetars have explosion energies $E_{\rm SNR}\lesssim10^{51}$ erg. Since the remnants associated with these magnetars are probably in the Sedov phase, thermal radiation loss from the remnants is nearly negligible, without involving other energy sources, we could assume $E_{\rm SNR} \simeq E_{\rm inj,s}$ (see also \cite{Vink:2006,Dall'Osso:2009}). Consequently, \textbf{for the slowly-spinning magnetars formed in weak SN explosions}, using the observed upper limit $E_{\rm SNR}\lesssim10^{51}$ erg and Eq. (\ref{Einj}), we can constrain their initial spin periods $P_{\rm i}$ \textbf{when other physical parameters of the magnetars at birth are determined}.

Based on the observations and theoretical work on the magnetic fields of magnetars introduced in Sec. \ref{Sec I}, the ratio of internal toroidal to surface dipole fields is taken to be $\bar{B}_{\rm t}/B_{\rm d}=2-100$. Furthermore, the maximum value of the toroidal field should not surpass the upper limit required by stable stratification, i.e., $\bar{B}_{\rm t}\leq 10^{17}$ G is required \cite{Reisenegger:2009}. These two conditions determine the reasonable range of $\bar{B}_{\rm t}$ of newborn magnetars. We take $T_{\rm c}=10^9$ K for the critical temperature and $T_{\rm i}=10^{10}$ K for the initial stellar temperature. The initial value for the tilt angle is set as $\chi_{\rm i}=1^{\circ}$. In Fig. \ref{Fig1}, we show the evolution of the \textbf{cumulative} energy injected into the ejecta by the newborn magnetar $E_{\rm inj}$ with time $t$. Here an ideal energy conversion efficiency $\eta=1$ is adopted. Panels (a)-(d) respectively correspond to $B_{\rm d}=10^{14}$, $5\times10^{14}$, $10^{15}$, and $5\times10^{15}$ G. These four values for $B_{\rm d}$ adopted approximately cover the typical strength of dipole fields of magnetars \cite{Olausen:2014}. To illustrate how the toroidal field can affect the results, in panels (a)-(c) we take $\bar{B}_{\rm t}=2B_{\rm d}$, $10B_{\rm d}$, and $100B_{\rm d}$, as indicated by the legends. In panel (d), the maximum toroidal field is limited to $\bar{B}_{\rm t}=10^{17}$ G. By requiring that the total injected energy is no more than the observed upper limit $E_{\rm SNR}\lesssim10^{51}$ erg on the explosion energies of some SN remnants introduced above, i.e., $E_{\rm inj,s}\lesssim10^{51}$ erg, we can obtain the minimum initial spin periods $P_{\rm i, min}$ of magnetars \textbf{formed in weak SN explosions}. The upper limit on the energy is shown by the black dotted line in each panel of Fig. \ref{Fig1}.

\begin{figure}
\begin{center}
\includegraphics[width=4.6 in, height=4.6 in,
clip]{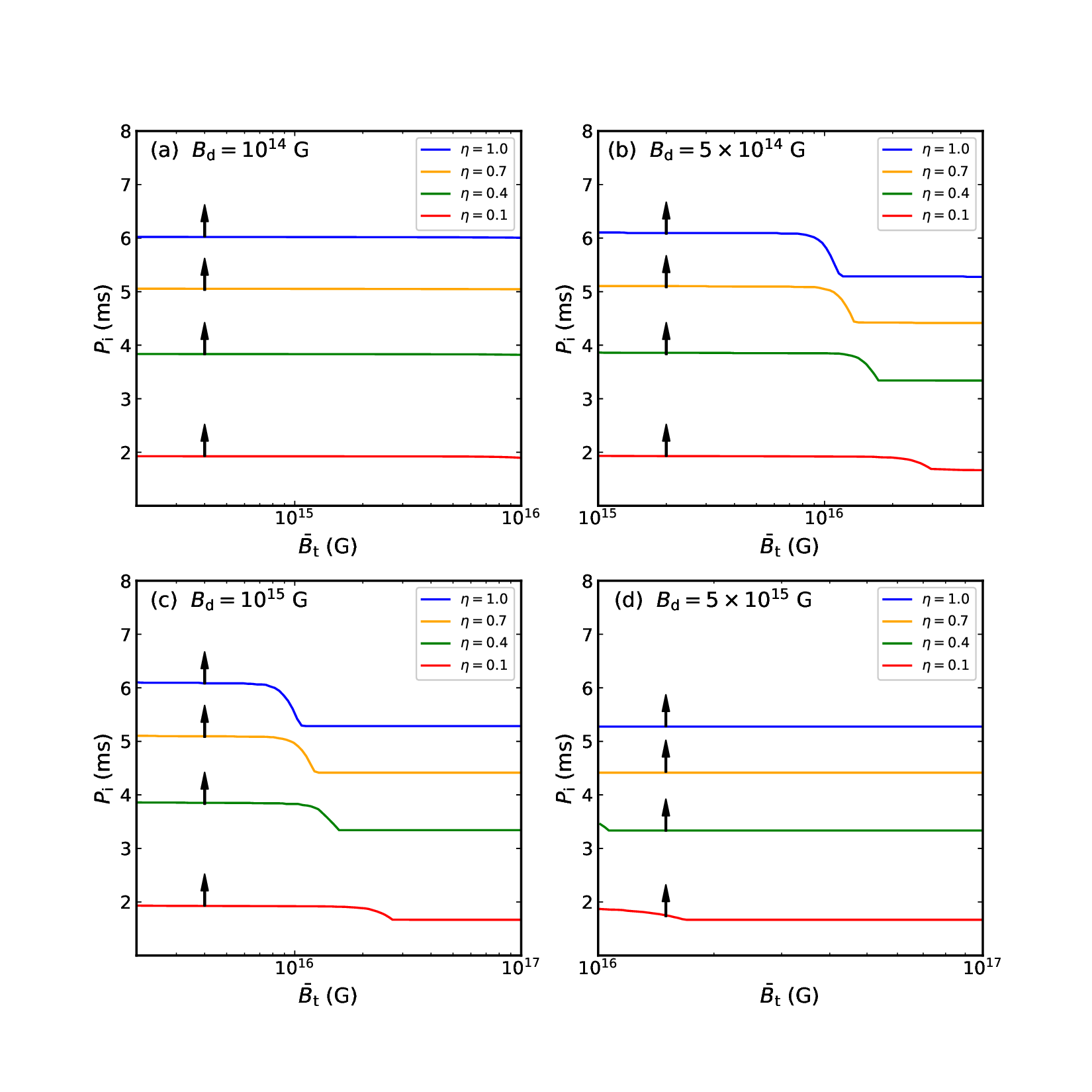} \caption{The parameter space of initial spin periods $P_{\rm i}$ versus toroidal fields $\bar{B}_{\rm t}$ of newborn magnetars \textbf{formed in weak SN explosions with $E_{\rm SNR}\lesssim10^{51}$ erg}. In panels (a)-(d), the dipole fields are respectively taken as $B_{\rm d}=10^{14}$, $5\times10^{14}$, $10^{15}$, and $5\times10^{15}$ G. We adopt a reasonable range $2B_{\rm d}\leq\bar{B}_{\rm t}\leq100B_{\rm d}$ for the toroidal fields in panels (a)-(c), while in panel (d), the maximum strength is set as $\bar{B}_{\rm t}=10^{17}$ G. The colored curves represent the minimum initial spin periods $P_{\rm i, min}$ of \textbf{these} newborn magnetars with $B_{\rm d}$ and $\bar{B}_{\rm t}$ adopted above, which are derived by using $E_{\rm SNR}\lesssim10^{51}$ erg and assuming specific values for $\eta$, as shown in the legends. } \label{Fig3}
\end{center}
\end{figure}

Obviously, after the birth of the newborn magnetar, the energy of MD radiation is gradually injected into the SN ejecta and finally $E_{\rm inj}$ becomes saturated at $t_{\rm sat}$, which depends on the strength of $B_{\rm d}$. Panel (a) shows that newborn magnetars with $B_{\rm d}=10^{14}$ G and reasonable $\bar{B}_{\rm
t}$ are allowed to have $P_{\rm i, min}\simeq 6$ ms if $\eta=1$ is assumed, irrespective of the specific values of $\bar{B}_{\rm t}$. Smaller $P_{\rm i}$ will lead to $E_{\rm inj,s}>10^{51}$ erg. Likewise, newborn magnetars with $B_{\rm d}=5\times10^{14}$ G may have $5.28\lesssim P_{\rm i, min}\lesssim 6.10$ ms for $\eta=1$, however, a larger $\bar{B}_{\rm t}$ will result in a smaller $P_{\rm
i, min}$ in this case, as presented in panel (b). The anti-correlation between $\bar{B}_{\rm t}$ and $P_{\rm i, min}$ can also be found in panel (c), which shows the results of magnetars with $B_{\rm d}=10^{15}$ G. The reason may be that as $\bar{B}_{\rm t}$ of newborn magnetars with $B_{\rm d}=5\times10^{14}$ and $10^{15}$ G increases, more spin energy of the magnetars could be lost via GW radiation, hence a smaller $P_{\rm i, min}$ is allowed. Panels (b) and (c) indicate that newborn magnetars with $B_{\rm d}=5\times10^{14}$ and $10^{15}$ G have the same allowed range for $P_{\rm i, min}$. Panel (d) shows that newborn magnetars with $B_{\rm d}=5\times10^{15}$ have $P_{\rm i, min}=5.28$ ms, also irrespective of the values of $\bar{B}_{\rm t}$. Overall, to satisfy the upper limit $E_{\rm SNR}\lesssim10^{51}$ erg on the explosion energies of SN remnants associated with \textbf{slowly-spinning magnetars born in weak SN explosions}, the minimum initial spin periods of these magnetars with dipole fields $10^{14}\leq B_{\rm d}\leq 5\times10^{15}$ G should be within $5.28\lesssim P_{\rm i, min}\lesssim 6.10$ ms when an ideal efficiency $\eta=1$ is assumed.

In Fig. \ref{Fig2}, the results for a possible conversion efficiency $\eta=0.4$ \cite{Woosley:2010} are presented for comparison. The magnetic fields (both $B_{\rm d}$ and $\bar{B}_{\rm t}$) are taken the same as in Fig. \ref{Fig1}. From Fig. \ref{Fig2} we find that for newborn magnetars with $10^{14}\leq B_{\rm d}\leq 5\times10^{15}$ G and reasonable strengths of $\bar{B}_{\rm t}$, their minimum initial spin periods are $3.34\lesssim P_{\rm i, min}\lesssim 3.86$ ms. Therefore, as $\eta$ decreases from 1 to 0.4, newborn magnetars are allowed to have smaller $P_{\rm i, min}$. Specifically, newborn magnetars with $B_{\rm d}=10^{14}$, $5\times10^{14}$, $10^{15}$, and $5\times10^{15}$ G respectively have $3.82\lesssim P_{\rm i, min}\lesssim 3.85$, $3.34\lesssim P_{\rm i, min}\lesssim 3.86$, $3.34\lesssim P_{\rm i, min}\lesssim 3.86$, and $3.34\lesssim P_{\rm i, min}\lesssim 3.49$ ms when $\eta=0.4$ is assumed.

The effect of $\eta$ on $P_{\rm i, min}$ of the newborn magnetars \textbf{formed in weak SN explosions} can be found in Fig. \ref{Fig3}, which shows the curve of $P_{\rm i, min}$ versus $\bar{B}_{\rm t}$ obtained by adopting different $\eta$ (see the legends). Panels (a)-(d) respectively correspond to $B_{\rm d}=10^{14}$, $5\times10^{14}$, $10^{15}$, and $5\times10^{15}$ G, while $\bar{B}_{\rm t}$ are confined by both $2B_{\rm d}\leq\bar{B}_{\rm t}\leq 100B_{\rm d}$ and $\bar{B}_{\rm t}\leq10^{17}$ G. Therefore, both $B_{\rm d}$ and $\bar{B}_{\rm t}$ are within reasonable ranges. The parameter spaces below these colored lines are excluded because they can lead to the violation of $E_{\rm SNR}\lesssim10^{51}$ erg. For \textbf{these} newborn magnetars, we can set constraints on their $P_{\rm i}$, though the constraints are $\eta$-dependent. Fig. \ref{Fig3} \textbf{shows that} with the decrease of $\eta$, \textbf{these} newborn magnetars are generally allowed to have smaller $P_{\rm i, min}$. Specifically, the newborn magnetars could have a very fast initial spin of $1<P_{\rm i, min}\lesssim 2$ ms if the conversion efficiency is as low as $\eta=0.1$. Our results thus differ from that of Dall'Osso et al. \cite{Dall'Osso:2009}, which suggested that $P_{\rm i}\sim1-2$ ms is allowed even for $\eta=1$. Panels (b) and (c) display that for a specific $\eta$ adopted, $P_{\rm i, min}$ nearly keeps unchanged first, then gradually decreases, and remains nearly unchanged again with the increase of $\bar{B}_{\rm t}$. The reason may be as follows. When $\bar{B}_{\rm t}$ is below a certain strength, GW radiation is gradually enhanced with the increase of $\bar{B}_{\rm t}$, \textbf{these} newborn magnetars are therefore allowed to have smaller $P_{\rm i, min}$. However, further increase of $\bar{B}_{\rm t}$ could remarkably suppress the growth of the tilt angle and also GW radiation from \textbf{these} newborn magnetars \cite{Cheng:2015,Cheng:2018}, whereas it could slightly affect the MD radiation. Therefore, further increase of $\bar{B}_{\rm t}$ does not result in smaller $P_{\rm i, min}$. Although $P_{\rm i, min}$ may decrease with the increase of $\bar{B}_{\rm t}$, the variation in $P_{\rm i, min}$ is generally small ($\lesssim1$ ms) for a constant $\eta$, as found in panels (b) and (c). Taken as a whole, the results in Fig. \ref{Fig3} indicate that \textbf{for a fixed $\eta$, these newborn magnetars generally have similar $P_{\rm i, min}$ when their $B_{\rm d}$ and $\bar{B}_{\rm t}$ have reasonable strengths as adopted here.} Specifically, the minimum initial spin periods of \textbf{these} newborn magnetars are $P_{\rm i, min}\simeq 1-2$ ms for $\eta=0.1$, whereas $P_{\rm i, min}\simeq 3-4$ ms for $\eta=0.4$, and $P_{\rm i, min}\simeq 5-6$ ms for $\eta=1$.

\section{GW radiation from newborn magnetars formed in weak SN explosions}\label{Sec IV}

Since we have set constraints on the initial spin periods of newborn magnetars \textbf{formed in weak SN explosions} by using the upper limit on the explosion energies of SN remnants associated with \textbf{slowly-spinning} magnetars, an estimate of GWs emitted by these newborn magnetars can therefore be made. This can
be realized by adopting specific values for $B_{\rm d}$, $\bar{B}_{\rm t}$ and $P_{\rm i}$ of the newborn magnetars, and then studying the evolution of magnetars. As mentioned in Sec. \ref{Sec II}, the growth of $\chi$ may be suppressed if the newborn
magnetar has strong enough $\bar{B}_{\rm t}$. In the case of a small $\chi$, GWs from magnetic deformation of the magnetar are emitted at frequencies of both $\nu_{\rm e,1}=\nu$ and $\nu_{\rm e,2}=2\nu$ \cite{Cheng:2015,Marassi:2011}, where $\nu=\Omega/2\pi$ is the star's spin frequency. The strain amplitudes of GWs emitted at
$\nu_{\rm e,1}$ and $\nu_{\rm e,2}$ are respectively given as \cite{Wette:2023}
\begin{eqnarray}\label{ht1}
   h_1(t)=\frac{8\pi^2GI\epsilon_{\rm B}\nu_{\rm e,1}^2}{c^4D}\sin(2\chi),
\end{eqnarray}
and
\begin{eqnarray}\label{ht2}
   h_2(t)=\frac{8\pi^2GI\epsilon_{\rm B}\nu_{\rm e,2}^2}{c^4D}\sin^2\chi,
\end{eqnarray}
where $D$ is the distance to the source. The characteristic amplitudes of GWs emitted at $\nu_{\rm e,1}$ and $\nu_{\rm e,2}$ are thus derived as $h_{\rm c}(\nu_{\rm e,1})=\nu_{\rm e,1}h_1(t)/\sqrt{d\nu_{\rm e,1}/dt}$ and $h_{\rm c}(\nu_{\rm e,2})=\nu_{\rm e,2}h_2(t)/\sqrt{d\nu_{\rm e,2}/dt}$, respectively.

\begin{figure}[h]
  \begin{minipage}{0.45\linewidth}
  \centering
   \includegraphics[width=66mm]{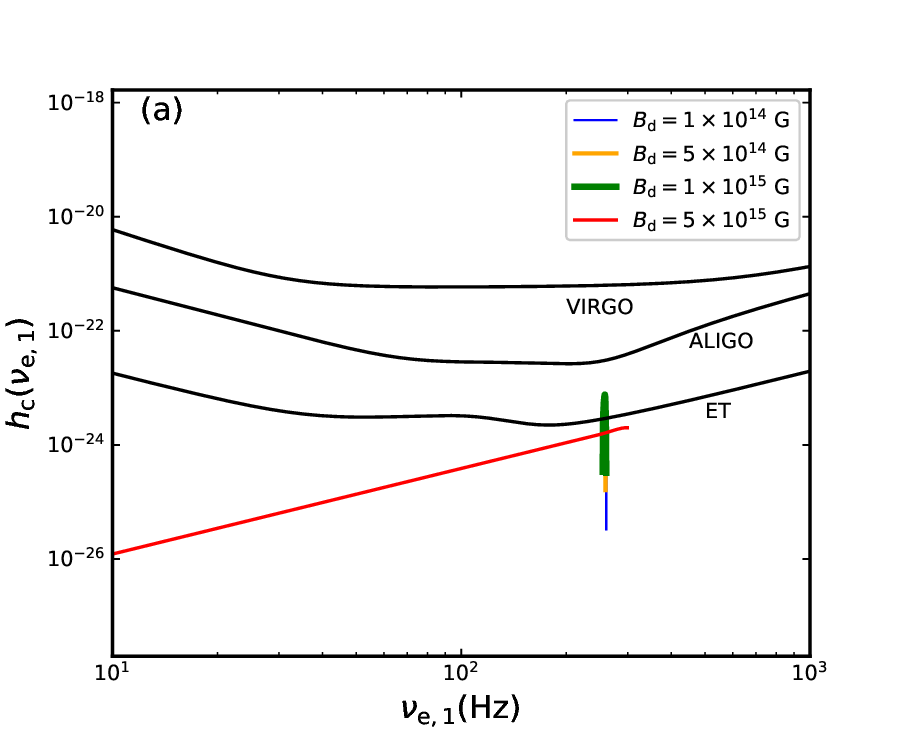}
  \end{minipage}
  \qquad
  \begin{minipage}{0.45\linewidth}
  \centering
   \includegraphics[width=66mm]{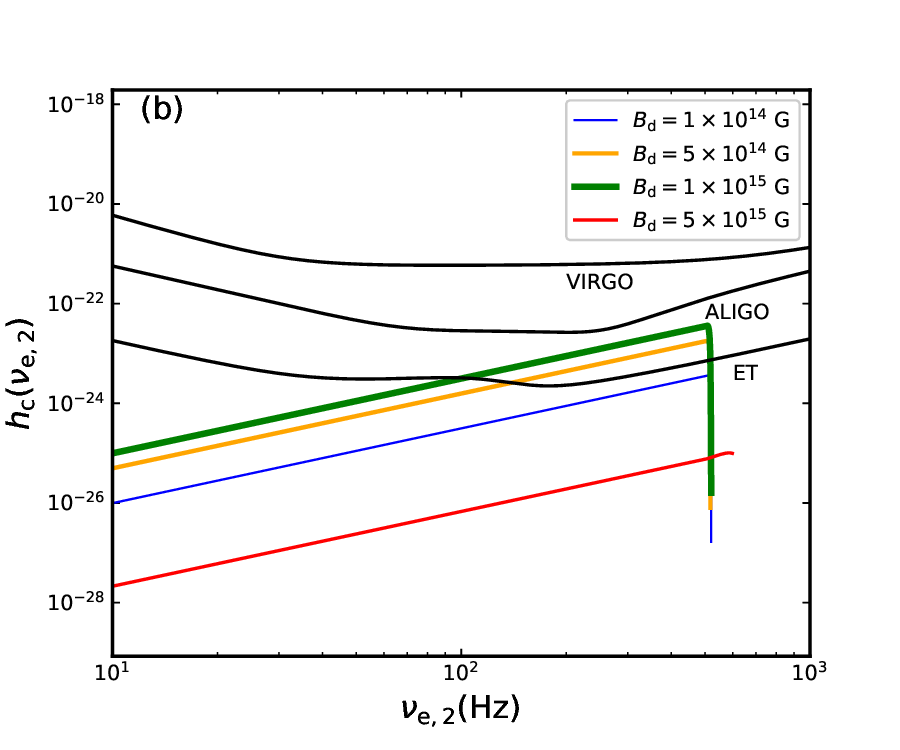}
  \end{minipage}%
    \caption{\label{Fig4}{\textbf{Panel (a)}: The characteristic amplitude $h_{\rm c}(\nu_{\rm e,1})$ versus emitted frequency $\nu_{\rm e,1}=\nu$ of GWs from newborn magnetars \textbf{formed in weak SN explosions}, where $\nu$ are the spin frequencies of the magnetars. Their dipole fields $B_{\rm d}$ are indicated in the legends. The black lines labeled respectively show the rms strain noises (e.g., \cite{Sathyaprakash:2009,Cheng:2017,Cheng:2023}) of the VIRGO, ALIGO at design sensitivity, and future ET. \textbf{Panel (b)}: The same as in panel (a), however, the curves of $h_{\rm c}(\nu_{\rm e,2})$ versus $\nu_{\rm e,2}=2\nu$ are shown here. Please see the text for details about the two panels.}}
\end{figure}

In panels (a) and (b) of Fig. \ref{Fig4}, the curves of $h_{\rm c}(\nu_{\rm e,1})$ versus $\nu_{\rm e,1}$ and $h_{\rm c}(\nu_{\rm e,2})$ versus $\nu_{\rm e,2}$ are respectively shown. \textbf{Following Dall'Osso et al. \cite{Dall'Osso:2009}, we take $D=20$ Mpc here, which represents the distance from the Virgo supercluster to the earth.} To intuitively show the amplitudes of the emitted GWs, the rms strain noises (see, e.g., \cite{Sathyaprakash:2009,Cheng:2017,Cheng:2023}) of the VIRGO, Advanced LIGO (ALIGO) at design sensitivity, and future Einstein Telescope (ET) are also presented for comparison. In the calculations, the dipole fields of the newborn magnetars are taken as $B_{\rm d}=10^{14}$, $5\times10^{14}$, $10^{15}$, and $5\times10^{15}$ G (see the legends), while the toroidal fields are adopted as $\bar{B}_{\rm t}=10B_{\rm d}$. Assuming a possible conversion efficiency $\eta=0.4$, from Fig. \ref{Fig3} we find that the newborn magnetars with $B_{\rm d}=10^{14}$, $5\times10^{14}$, $10^{15}$, and $5\times10^{15}$ G respectively have $P_{\rm i,min}=3.84$, 3.86, 3.84, and 3.34 ms. For simplicity, here we assume $P_{\rm i}=P_{\rm i,min}$ because this can result in the largest amplitudes and highest frequencies of GWs.

Panel (a) of Fig. \ref{Fig4} shows that for the physical parameters ($B_{\rm d}$, $\bar{B}_{\rm t}$, and $P_{\rm i}$) of newborn magnetars considered here, the GWs emitted at $\nu_{\rm e,1}$ by these magnetars have very small characteristic amplitudes, \textbf{which seem to be hardly detectable even by the ET}. Moreover, the tilt angles of newborn magnetars with $B_{\rm d}=10^{14}$, $5\times10^{14}$, and $10^{15}$ G can increase to $\pi/2$ in a very short time, during which these magnetars barely spin down. Consequently, the GWs emitted at $\nu_{\rm e,1}$ by these magnetars
are almost monochromatic with frequency centering at $\sim260$ Hz, as indicated by the blue, yellow, and green lines. In contrast, the tilt angle of the newborn magnetar with $B_{\rm d}=5\times10^{15}$ G remains to be small for a long time and increases to $\pi/2$ in the second stage. During this process, the spin frequency of this magnetar has decreased significantly. Therefore, though the GWs emitted at $\nu_{\rm e,1}$ by this magnetar covers a wide frequency range (see the red line), $h_{\rm c}(\nu_{\rm e,1})$ is actually rather small because a small $\chi$ is maintained for a long time.
For the same reason, the GW radiation at $\nu_{\rm e,2}$ from this newborn magnetar is also suppressed and probably undetectable even by the ET, as shown in panel (b) of Fig. \ref{Fig4}. However, the newborn magnetars with $B_{\rm d}=5\times10^{14}$ and $10^{15}$ G have strong enough magnetic fields and their $\chi$ can increase to $\pi/2$ in a very short time, the GWs emitted at $\nu_{\rm e,2}$ from these magnetars \textbf{have relatively large amplitudes}. Furthermore, the highest frequency of the GWs detected possibly reaches $\sim500$ Hz when a possible conversion efficiency $\eta=0.4$ is adopted, and may be as high as $\sim1000$ Hz if the efficiency is as low as $\eta=0.1$. Consequently, to detect the GWs from magnetic deformation of \textbf{the newborn magnetars formed in weak SN explosions}, one may need to mainly focus on the frequency domain $\lesssim 500$ Hz, as inferred from the constraints on the initial spin periods of \textbf{these} magnetars.

\textbf{We also make a quantitative analysis about the detection of the GWs mentioned above by calculating the optimal\footnote{This can be realized by using the method of matched filter in the detection.} signal-to-noise ratio (${\rm S/N}$) of the GW signals for a ground-based detector. The specific form of the optimal ${\rm S/N}$ is expressed as \cite{Cheng:2017,Corsi:2009}}
\begin{eqnarray}\label{snr}
    {\rm S/N}=\sqrt{\int_{\nu_{\rm e, min}}^{\nu_{\rm e, max}}\frac{h_{\rm c}^2d\nu_{\rm e}}{\nu_{\rm e}^2S_h(\nu_{\rm e})}},
\end{eqnarray}
\textbf{where $\nu_{\rm e, min}$ and $\nu_{\rm e, max}$ are the minimum and maximum frequencies of GWs emitted by the magnetar. $S_h(\nu_{\rm e})$ represents the detector's one-sided noise power spectral density \cite{Sathyaprakash:2009}.}

\begin{figure}[h]
  \begin{minipage}{0.45\linewidth}
  \centering
   \includegraphics[width=66mm]{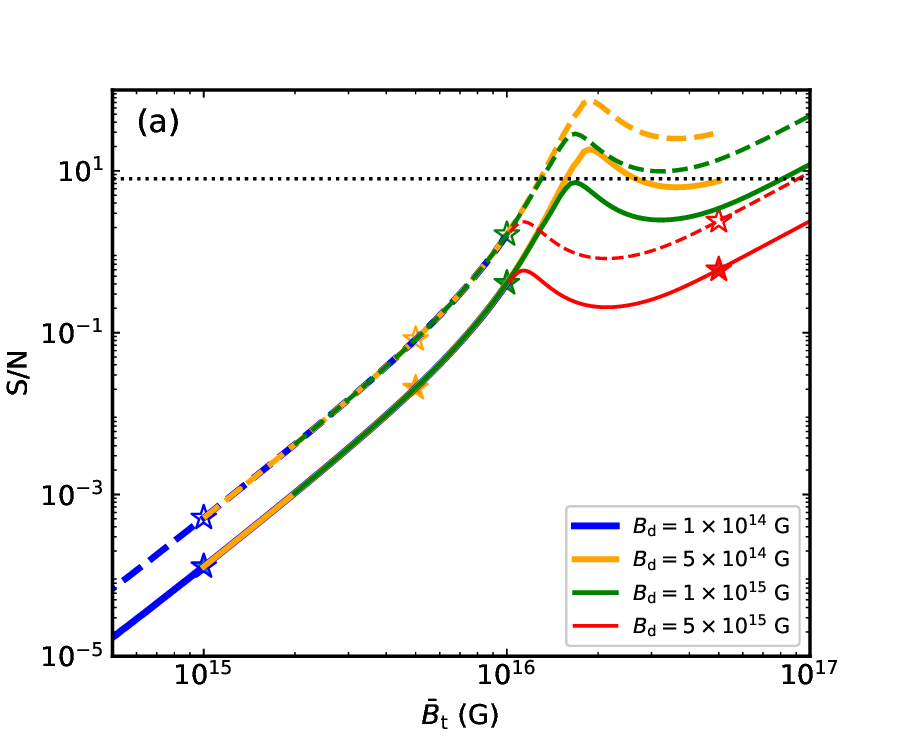}
  \end{minipage}
  \qquad
  \begin{minipage}{0.45\linewidth}
  \centering
   \includegraphics[width=66mm]{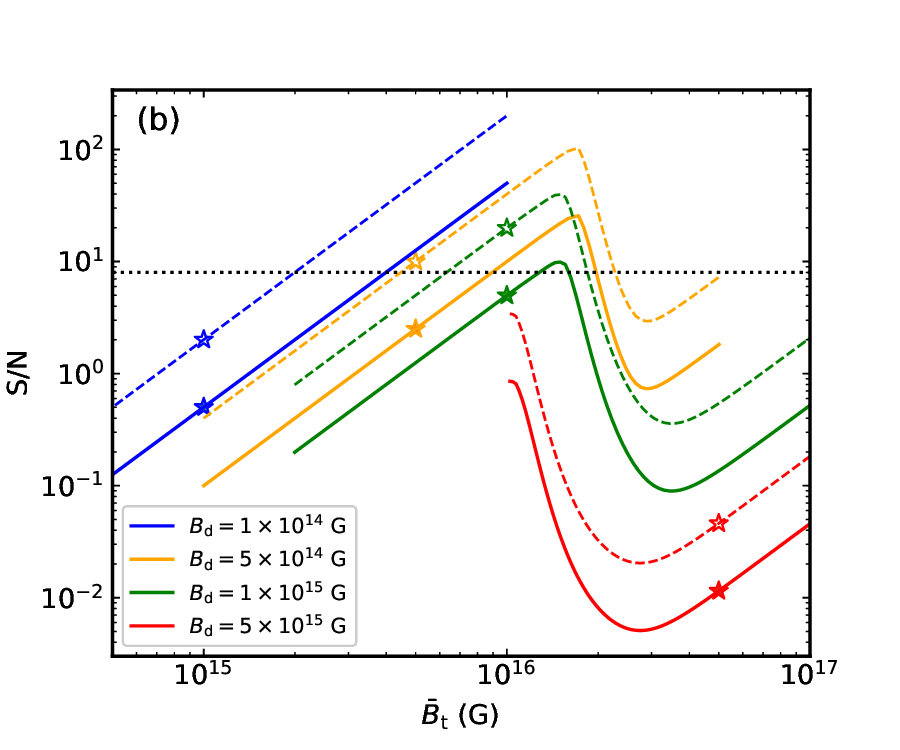}
  \end{minipage}%
  \caption{\label{Fig5}{\textbf{The optimal ${\rm S/N}$ of the GW signals emitted at $\nu_{\rm e,1}$ (Panel (a)) and $\nu_{\rm e,2}$ (Panel (b)) versus toroidal field $\bar{B}_{\rm t}$ of newborn magnetars formed in weak SN explosions. Their dipole fields $B_{\rm d}$ are indicated in the legends. All ${\rm S/N}$ of the GWs are calculated regarding to the ET. The colored solid and dashed lines represent ${\rm S/N}$ derived by adopting $D=20$ and $5$ Mpc, respectively. ${\rm S/N}$ of the GW signals investigated in Fig. \ref{Fig4} are shown by the colored solid stars. Assuming these newborn magnetars are located at $5$ Mpc, ${\rm S/N}$ of the emitted GWs are indicated by the colored hollow stars. See the text for details.}}}
\end{figure}

\textbf{Fig. \ref{Fig5} shows the curves of the optimal ${\rm S/N}$ regarding to the ET versus $\bar{B}_{\rm t}$ of newborn magnetars formed in weak SN explosions. Panels (a) and (b) repectively present ${\rm S/N}$ of the GWs emitted at $\nu_{\rm e,1}$ and $\nu_{\rm e,2}$ by these magnetars. The dipole fields $B_{\rm d}$ of these magnetars are indicated in the legends, while the ranges of $\bar{B}_{\rm t}$ are taken the same as in Fig. \ref{Fig3}. The solid and dashed curves are obtained by using $D=20$ and $5$ Mpc, respectively. The colored solid stars in Panels (a) and (b) give ${\rm S/N}$ of the GW signals investigated in Fig. \ref{Fig4} by assuming the sources are located at $D=20$ Mpc. For comparison, the colored hollow stars show ${\rm S/N}$ of the GWs emitted by the newborn magnetars with the same physical parameters, however, are located at a very close distance of $5$ Mpc instead. Panel (a) shows that for the reasonable values of the physical parameters adopted, even if the newborn magnetars are located at $5$ Mpc, the GWs emitted at $\nu_{\rm e,1}$ from these magnetars have very low ${\rm S/N}$. The maximum signal-to-noise ratio is ${\rm S/N}_{\rm max}=2.41$ (the red hollow star), which is below the detection threshold (${\rm S/N}_{\rm th}=8$ as depicted by the black dotted lines in the two panels) of a single-detector search \cite{Cheng:2017,Abadie:2010}. Therefore the GWs emitted at $\nu_{\rm e,1}$ from these newborn magnetars are probably undetectable even by the ET.}

\textbf{Adopting reasonable values for the physical parameters of these newborn magnetars, the GWs emitted at $\nu_{\rm e,2}$ in some cases have relatively large signal-to-noise ratios with ${\rm S/N}=10.01$ and 19.85 (the yellow hollow and green hollow stars in panel (b)), both are above the detection threshold ${\rm S/N}_{\rm th}=8$. This suggests that the GWs emitted at $\nu_{\rm e,2}$ by the newborn magnetars with $B_{\rm d}=5\times10^{14}$ and $10^{15}$ G may be detectable for the ET if these magnetars are located at $D=5$ Mpc. However, for a larger distance of $D=20$ Mpc, the GWs emitted at $\nu_{\rm e,2}$ by these newborn magnetars are possibly undetectable for the ET because the maximum ${\rm S/N}$ is only 4.96 (the green solid star in panel (b)). Finally, the resluts in Fig. \ref{Fig5} show that ${\rm S/N}$ of the GWs emitted at both $\nu_{\rm e,1}$ and $\nu_{\rm e,2}$ do not increase monotonically with the increase $\bar{B}_{\rm t}$ for the newborn magnetars with $B_{\rm d}=5\times10^{14}$, $10^{15}$, and $5\times10^{15}$ G. The reason may be as follows. When $\bar{B}_{\rm t}$ of these magnetars are large enough, further increase of $\bar{B}_{\rm t}$ could strongly suppress the growth of $\chi$, leading to the suppression of GW radiation from these magnetars. In contrast, as $\bar{B}_{\rm t}$ increases, $\chi$ of the newborn magnetar with $B_{\rm d}=10^{14}$ G can always increase to $\pi/2$ in a very short time, thus its GW radiation is not suppressed.}  

\section{Conclusion and Discussions}\label{Sec V}

In this work, we have set constraints on the initial spin periods $P_{\rm i}$ of newborn magnetars \textbf{formed in weak SN explosions} by using the upper limit $E_{\rm SNR}\lesssim10^{51}$ erg on the explosion energies of the SN remnants around \textbf{slowly-spinning} magnetars. This upper limit was used in \cite{Dall'Osso:2009} to constrain the parameter space of magnetic fields of \textbf{these} newborn magnetars. Although our method is generally the same as that of \cite{Dall'Osso:2009}, there are actually some improvements. The main improvement is that we considered both the first and the second stages of the tilt angle evolution of \textbf{these} newborn magnetars, while only the first stage of evolution was involved in \cite{Dall'Osso:2009}. The second stage of tilt angle evolution is important, especially for the magnetars with large enough toroidal fields, thus should be taken into account when studying the evolution of \textbf{these} newborn magnetars. Second, following previous observational results and theoretical work on the magnetic fields of magnetars, the toroidal fields here are required to satisfy $2B_{\rm d}\leq\bar{B}_{\rm t}\leq100B_{\rm d}$ and $\bar{B}_{\rm t}\leq10^{17}$ G, rather than allowing them to have unreasonable large values. Third, we involved the conversion efficiency $\eta$ of the EM energy from MD radiation into the kinetic energy of the ejecta and treated it as a free parameter since its value is highly uncertain.

Our results show that the minimum initial spin periods $P_{\rm i,min}$ of \textbf{these} newborn magnetars are mainly $\eta$-dependent, however, slightly affected by $B_{\rm d}$ and $\bar{B}_{\rm t}$ if they have reasonable strengths as considered in this work. We find that an ideal efficiency $\eta=1$ generally corresponds to $P_{\rm i,min}\simeq 5-6$ ms, while a possible efficiency $\eta=0.4$ \cite{Zhou:2019,Woosley:2010} leads to $P_{\rm i, min}\simeq 3-4$ ms. When the efficiency is as low as $\eta=0.1$, we have $P_{\rm i,min}\simeq 1-2$ ms. In contrast, as found in \cite{Dall'Osso:2009}, \textbf{these} newborn magnetars are allowed to have $P_{\rm i}\sim1-2$ ms even if $\eta=1$ is adopted. Using these constraints, we also estimated the characteristic amplitudes of GWs emitted by these magnetars \textbf{and the corresponding ${\rm S/N}$ for the ET. Assuming typical values for $B_{\rm d}$, $\bar{B}_{\rm t}=10B_{\rm d}$, $\eta=0.4$, and $P_{\rm i}=P_{\rm i, min}$, the GWs emitted at both $\nu_{\rm e,1}=\nu$ and $\nu_{\rm e,2}=2\nu$ may be undetectable for the ET if the newborn magnetars located at $20$ Mpc away in the Virgo supercluster because their ${\rm S/N}$ are all below 8. Assuming a closer distance of $5$ Mpc, only the GWs emitted at $\nu_{\rm e,2}=2\nu$ from the newborn magnetars with $B_{\rm d}=5\times10^{14}$ and $10^{15}$ G could be detected by the ET since the signals respectively have ${\rm S/N}=10.01$ and 19.85}. Moreover, a possible conversion efficiency $\eta=0.4$ indicates that the potentially detectable GWs from these newborn magnetars are emitted in the frequency domain $\lesssim500$ Hz. This may help to narrow down the frequency range when searching for GWs from these newborn magnetars. 

\textbf{As is well known, core-collapse SNe that produce NSs can also emit GWs, and the signals may be detected by the ground-based detectors if the sources are close enough to the earth \cite{Ott:2009}. The numerical simulations of magnetohydrodynamically-driven core-collapse SNe (that could produce magnetars) performed by Takiwaki \& Kotake \cite{Takiwaki:2011} showed that the characteristic amplitudes of the emitted GWs in the frequency range $\sim100-500$ Hz are $h_{\rm c}\sim10^{-21}-10^{-20}$ if the sources are located at 10 kpc away (see also \cite{Scheidegger:2010}). These correspond to $h_{\rm c}\sim5\times10^{-25}-5\times10^{-24}$ if the SNe are located at 20 Mpc away. Comparing their results with panel (b) of Fig. \ref{Fig4}, we find that $h_{\rm c}$ of the GWs in $\sim100-500$ Hz from magnetohydrodynamically-driven core-collapse SNe are at most comparable to (and generally smaller than) that of the GWs emitted at $\nu_{\rm e,2}$ by the newborn magnetars with $B_{\rm d}=5\times10^{14}$ and $10^{15}$ G. In view of this, detection of the GWs from magnetic deformation of the newborn magnetars seems to be easier if the magnetars have $\epsilon_{\rm B}\sim10^{-4}$ (for $\bar{B}_{\rm t}=10^{16}$ G). However, the direct search for GWs from NSs in young SN remnants using data from the first half of the third observing run of ALIGO and advanced VIRGO showed no evidence of GWs, suggesting that the ellipticities of these NSs should be $<10^{-6}$ when the frequencies of GWs are $\gtrsim100$ Hz \cite{Abbott:2021}. In fact, the limit on the ellipticities of these NSs is model dependent, and the equation of state, the moment of inertia, and the magnetic fields can all affect the final results \cite{Wette:2023,Abbott:2021}. As a result, the direct search for GWs from NSs in young SN remnants cannot rule out the possibility that the newborn magnetars formed in weak SN explosions may have relatively large $\epsilon_{\rm B}$, especially when considering that the two kinds of NSs may have totally different magnetic fields}. 

The constraints on $P_{\rm i}$ of newborn magnetars \textbf{formed in weak SN explosions} may also shed light on the origin of \textbf{their} strong magnetic fields. Our results suggest that without violating the upper limit $E_{\rm SNR}\lesssim10^{51}$ erg, \textbf{these} newborn magnetars are allowed to have $P_{\rm i, min}\simeq 1-2$ ms \textbf{only when} the conversion efficiency is as low as $\eta=0.1$. \textbf{However, observational evidence supporting such a low efficiency is still lacking currently. Theoretically, in some energetic SN explosions, for instance, hypernovae \cite{Woosley:2006} associated with long GRBs, $\eta$ could be small because in this case newborn magnetars may have initial spin periods of $\sim 1-2$ ms \cite{Zhang:2001}, and thus GW emissions may be considerably amplified. Assuming a possible efficiency $\eta=0.4$ \cite{Zhou:2019,Woosley:2010}, the newborn magnetars formed in weak SN explosions have $P_{\rm i, min}\simeq3-4$ ms, suggesting that they are not very rapidly rotating ($\gg1$ ms) at birth. Actually, magnetar-strength magnetic fields could be produced because of the convective dynamo in nascent NSs even though they have a relatively slow initial spin of several milliseconds \cite{Raynaud:2020}}. Therefore, the upper limit $E_{\rm SNR}\lesssim10^{51}$ erg in principle could not exclude the dynamo origin of strong magnetic fields of magnetars \textbf{formed in weak SN explosions}.

\textbf{Finally, the constraints on $P_{\rm i}$ derived here probably cannot be directly applied to the newborn magnetars formed in more energetic SN explosions, e.g., hypernovae \cite{Woosley:2006} and SLSNe \cite{Kasen:2010}. To set constraints on $P_{\rm i}$ of these newborn magnetars, changes to the analytic model of magnetar evolution are required given that neutrino emissions possibly play an important role in the evolution. Furthermore, X-ray/radio observations of the remnants of hypernovae and SLSNe that may harbor magnetars are also necessary in order to determine the explosion energies of these remnants. }

\begin{acknowledgements}
We thank the anonymous referee for helpful comments and suggestions, which helped improve the manuscript a lot. This work is supported by the National SKA program of China (Grant No. 2020SKA0120300), and the National Natural Science Foundation of China (Grant No. 12003009, and No. 12033001).
\end{acknowledgements}

\noindent\textbf{Data Availability Statement} 
This manuscript has no associated data. [Authors' comment: This is a theoretical study and no experimental data has been presented.]

\noindent\textbf{Code Availability Statement} 
This manuscript has no associated code/software. [Authors' comment: Code/Software sharing not applicable to this article as no code/software was generated or analyzed during the current study.]

\label{lastpage}
\end{document}